\documentclass[final,3p,times,sort&compress]{elsarticle}
\usepackage{multirow,setspace,times,amssymb,amsmath,graphicx,color,rotating,subfigure,url}
\usepackage{lineno}
\usepackage{natbib}
\makeatletter

\journal{Physica A} 

\begin{document}

\begin{frontmatter}

\title{On the growth of primary industry and population of China's counties}

\author[SB,SS]{Wen-Jie Xie}
\author[SB,SS]{Gao-Feng Gu}
\author[SB,SS,RCE]{Wei-Xing Zhou\corref{cor}}
\cortext[cor]{Corresponding author. Address: 130 Meilong Road, P.O.
Box 114, School of Business, East China University of Science and
Technology, Shanghai 200237, China, Phone: +86 21 64253634, Fax: +86
21 64253152.}
\ead{wxzhou@ecust.edu.cn} %

\address[SB]{School of Business, East China University of Science and Technology, Shanghai 200237, China}
\address[SS]{School of Science, East China University of Science and Technology, Shanghai 200237, China}
\address[RCE]{Research Center for Econophysics, East China University of Science and Technology, Shanghai 200237, China}

\begin{abstract}
The growth dynamics of complex organizations have attracted much interest of econophysicists and sociophysicists in recent years. However, most of the studies are done for developed countries. We investigate the growth dynamics of the primary industry and the population of 2079 counties in mainland China using the data from the China County Statistical Yearbooks from 2000 to 2006. We find that the annual growth rates are distributed according to Student's $t$ distribution with the tail exponent less than 2. We find power-law relationships between the sample standard deviation of the growth rates and the initial size. The scaling exponent is less than 0.5 for the primary industry and close to 0.5 for the population.
\end{abstract}


\end{frontmatter}

\section{Introduction}

The law of proportionate growth, also known as the law of proportionate effect or Gibrat's law, is an important fundamental regularity in industry economics \cite{Coase-1993}. Numerous studies have been conducted by economists and econophysicists on the growth dynamics of firms in a long history. Denoting $S(t)$ the size of a firm at time $t$, its logarithmic growth rate is defined as
\begin{equation}
 r(t) = \ln S(t)-\ln S(t-1).
 \label{Eq:rt}
\end{equation}
The basic assumptions of Gibrat's law state that (1) the growth rate $r(t+1)$ of a firm is independent of its size $S(t)$, (2) the successive growth rates $r(t)$ are temporally uncorrelated, and (3) the growth rate distribution $f(r)$ is Gaussian. It follows immediately that the firm size distribution is log-normal.

However, there is undoubtable empirical evidence showing that Gibrat's law does not hold. Using data of all publicly traded USA manufacturing firms within the years 1974-1993, it is found that the distribution $f(r|S)$ of the annual growth rates of firms with approximately the same size $S$ has an exponential form
\begin{equation}
    f(r|S) \triangleq f(r(t)|S(t-1))=\frac{1}{\sqrt{2}\sigma(S)}\exp\left(-\frac{\sqrt{2}|r-\overline{r}(S)|}{\sigma(S)}\right),
    \label{Eq:Pr:r:Exp}
\end{equation}
and its standard deviation $\sigma(S)$ scales as a power law with the firm size $S$
\begin{equation}
    \sigma(S) \sim S^{-\beta},
    \label{Eq:sigma:r:S}
\end{equation}
where the scaling exponent $\beta$ is about 0.20 \cite{Stanley-Amaral-Buldyrev-Havlin-Leschhorn-Maass-Salinger-Stanley-1996-Nature,Amaral-Buldyrev-Havlin-Leschhorn-Maass-Salinger-Stanley-Stanley-1997-JPIF}, where the exponent $\beta$ is related to the fluctuation exponent $H$ by $\beta=1-H$ \cite{Rybski-Buldyrev-Havlin-Liljeros-Makse-2009-PNAS}.
Further investigation finds that the growth rates are distributed exponentially in the bulk followed by power-law tails
\cite{Fu-Pammolli-Buldyrev-Riccaboni-Matia-Yamasaki-Stanley-2005-PNAS,Pammolli-Fu-Buldyrev-Riccaboni-Matia-Yamasaki-Stanley-2007-EPJB}
\begin{equation}
    f(r|S) \sim |r|^{-\zeta},
    \label{Eq:Pr:r:PL}
\end{equation}
where $\zeta\approx3$.

Very similar statistical regularities of the growth rates have been unveiled in diverse fields including other business firms \cite{Takayasu-Okuyama-1998-Fractals,Bottazzi-Dosi-Lippi-Pammolli-Riccaboni-2001-IJIO,Fabritiis-Pammolli-Riccaboni-2003-PA,Matia-Fu-Buldyrev-Pammolli-Riccaboni-Stanley-2004-EPL}, gross domestic product (GDP) \cite{Canning-Amaral-Lee-Meyer-Stanley-1998-EL,Lee-Amaral-Canning-Meyer-Stanley-1998-PRL,Podobnik-Horvatic-Pammolli-Wang-Stanley-Grosse-2008-PRE},
scientific output \cite{Plerou-Amaral-Gopikrishnan-Meyer-Stanley-1999-Nature,Matia-Amaral-Luwel-Moed-Stanley-2005-JASIST}, human and bird population \cite{Keitt-Stanley-1998-Nature,Matia-Amaral-Luwel-Moed-Stanley-2005-JASIST,Rozenfeld-Rybski-Andrade-Batty-Stanley-Makse-2008-PNAS}, research and development expenditure \cite{Plerou-Amaral-Gopikrishnan-Meyer-Stanley-1999-Nature}, and so on. The growth dynamics of these complex systems share striking similarity indicating that their underlying interactions of subunits might exhibit universal behaviors. Distinct numerical and analytic models have been proposed to understand the growth dynamics of complex organizations \cite{Buldyrev-Amaral-Havlin-Leschhorn-Maass-Salinger-Stanley-Stanley-1997-JPIF,Takayasu-Okuyama-1998-Fractals,Amaral-Buldyrev-Havlin-Salinger-Stanley-1998-PRL,Drossel-1999-PRL,Amaral-Gopikrishnan-Plerou-Stanley-2001-PA,Bottazzi-Secchi-2003-EL,Sutton-2003-PA,Wyart-Bouchaud-2003-PA,Mizuno-Takayasu-Takayasu-2004-PA,Fu-Pammolli-Buldyrev-Riccaboni-Matia-Yamasaki-Stanley-2005-PNAS,Fu-Buldyrev-Salinger-Stanley-2006-PRE,Yamasaki-Matia-Buldyrev-Fu-Pammolli-Riccaboni-Stanley-2006-PRE,Buldyrev-Pammolli-Riccaboni-Yamasaki-Fu-Matia-Stanley-2007-EPJB,Buldyrev-Growiec-Pammolli-Riccaboni-Stanley-2007-JEEA,Podobnik-Horvatic-Pammolli-Wang-Stanley-Grosse-2008-PRE,Riccaboni-Pammolli-Buldyrev-Ponta-Stanley-2008-PNAS}.

In this work, we investigate the annual growth rate distributions of the primary industry and the population of 2079 counties in mainland China within seven years, which are retrieved from the China County Statistical Yearbook from 2000 to 2006. The data sets are briefly described in Section~\ref{S1:data}. Section \ref{S1:PDF} studies the growth rate distributions of the two variables and Section~\ref{S1:Std:Size} studies the size dependence behavior of standard deviations of the growth rates. Section \ref{S1:Conclusion} summarizes.

\section{Data sets}
\label{S1:data}

The reform and opening-up policy has greatly affected China's economy and people's living standards since 1978. China has made great achievements in the economic development in the past 30 years with the average annual growth rate of GDP being about 9.8\%, which is much higher than the world average annual growth rate around 3.3\% in the same period. Compared with the beginning of the reform and opening-up, the GDP per capita is more than 2.4 thousand U.S. dollars with an annual growth of 8.2\%. It is thus interesting to investigate the growth dynamics of socio-economic variables of China.

The population and primary industry data analyzed in this work were collected from the China County Statistical Yearbooks from 2000 to 2006. The yearbooks were edited by the National Bureau of Statistics for the rural socio-economic survey and published by China Statistical Publishing House, which record the statistical information of the socio-economic development of the counties in mainland China, including economy, agriculture, industry, investment, education, health, and so on. The information in the yearbook covers all the counties except those in Hong Kong SAR, Macao SAR, and Taiwan Province. There are 2079 counties in total. In the yearbook, the total annual population represents the number of people in a certain period and in a certain area. The end annual population refers to the number of people just before December 31 each year in a certain area. Primary industry is the sum of agriculture, forestry, animal husbandry and fishery.

The basic statistics of the data sets are presented in Table~\ref{TB:DataStats}. We find that the values of kurtosis are all greatly larger than 3 for both the primary industry and the population from the year 2000 to 2006, which implies that the distributions have fat tails. On the other hand, the values of skewness are all negative which means that the distributions are asymmetric and skewed to the left. We standardize the growth rates to make $r$ have zero mean and unit variance.

\begin{table}[htp]
\centering
\caption{Basic statistics of the growth rates of the primary industry and the population of 2079 counties in mainland China.}
\medskip
\label{TB:DataStats}
\begin{tabular}{ccccccccccccccc}
\hline \hline
    \multirow{4}*[3.2mm] {Year} && \multicolumn{6}{c}{Primary Industry} && \multicolumn{6}{c}{Population} \\
    \cline{3-8} \cline{10-15}
    && Max & Min & Mean & Std & Skew & Kurt && Max & Min & Mean & Std & Skew & Kurt \\
    \hline
        2001&& 0.94& -3.52& 0.01& 0.22& -6.29& 77.19&& 0.94& -1.41& 0.00& 0.08& -4.44& 153.76\\
        2002&& 1.07& -2.26& 0.06& 0.17& -1.60& 48.98&& 0.34& -0.51& 0.00& 0.04& -1.13& 38.62\\
        2003&& 1.06& -2.27& 0.05& 0.17& -2.53& 33.10&& 0.71& -2.40& 0.00& 0.09& -12.76& 333.23\\
        2004&& 1.08& -5.91& 0.14& 0.38& -11.21& 156.39&& 0.92& -0.76& 0.01& 0.06& 2.09& 114.95\\
        2005&& 0.91& -2.29& 0.08& 0.18& -5.99& 77.06&& 0.65& -2.51& -0.00& 0.10& -17.45& 410.42\\
        2006&& 0.90& -6.54& 0.07& 0.19& -25.23& 871.57&& 0.69& -2.40& 0.00& 0.09& -15.24& 348.97\\
\hline\hline
\end{tabular}
\end{table}

\section{Probability distributions of the standardized growth rates}
\label{S1:PDF}

\subsection{Unconditional distributions}
\label{S2:PDF:unCond}

\begin{figure}[htb]
\centering
\includegraphics[width=8cm]{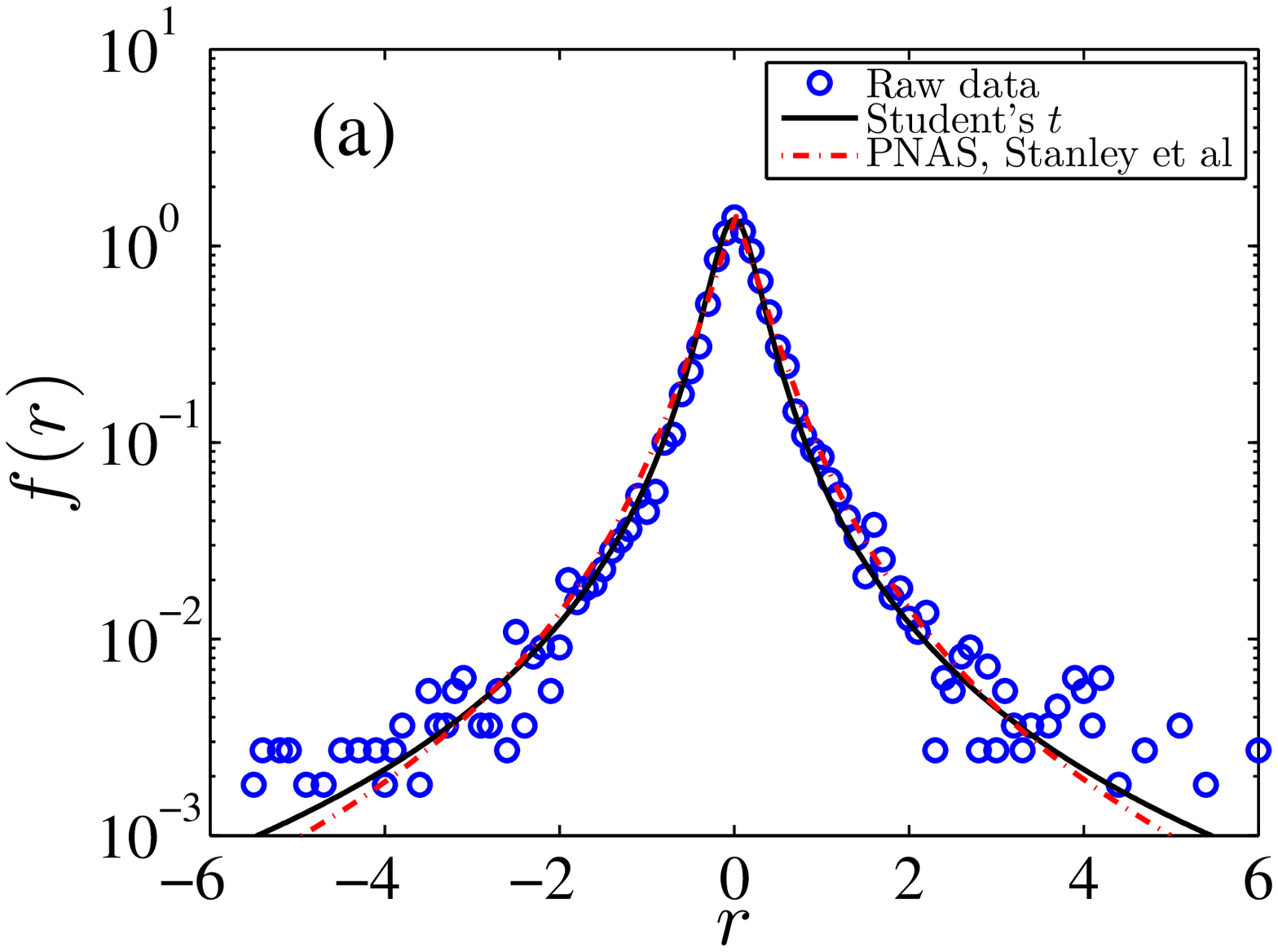}
\includegraphics[width=8cm]{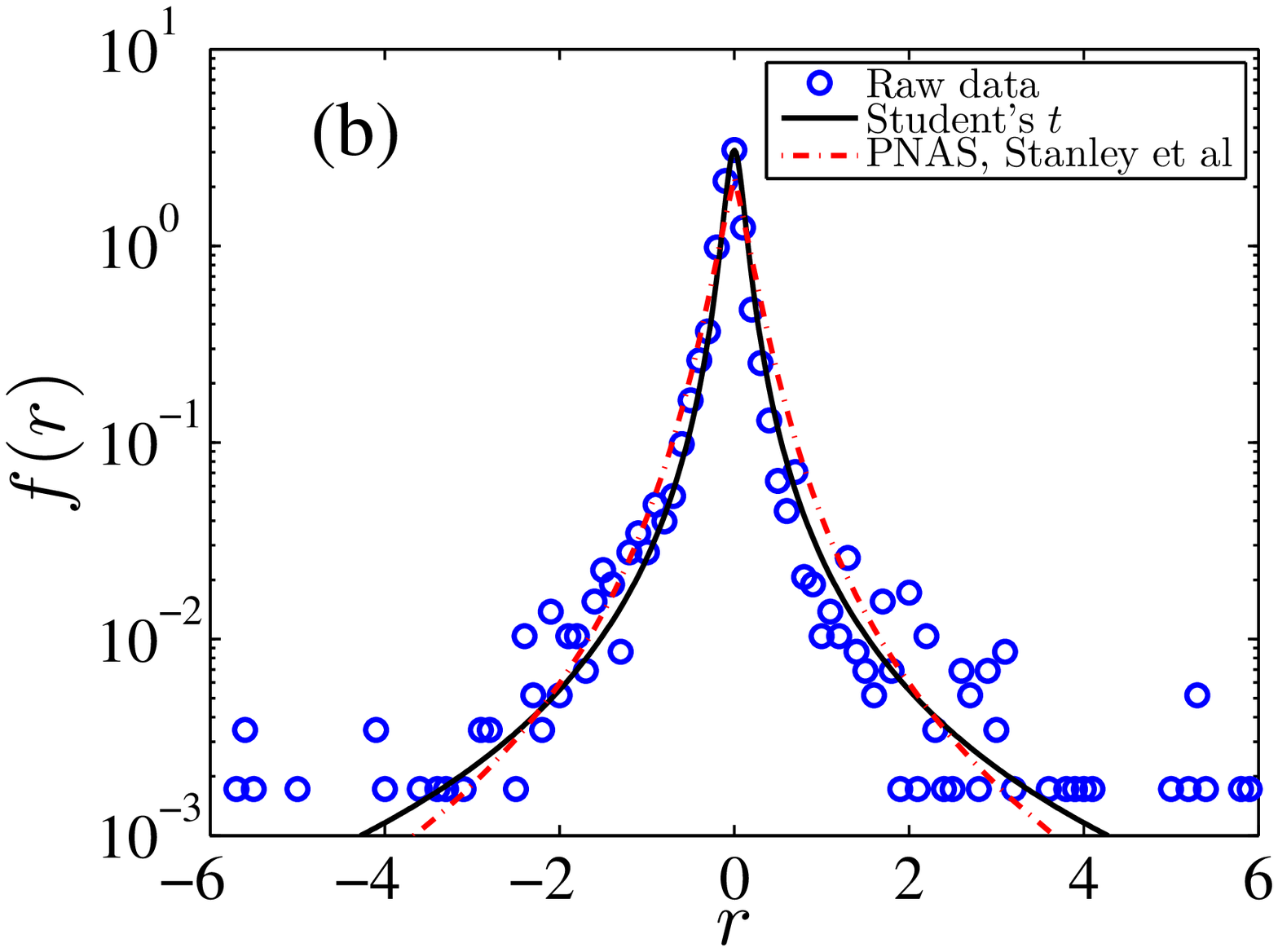}
\caption{\label{Fig:PDF} Empirical probability density functions $f(r)$ of the standardized annual growth rates $r$ for the primary industry (a) and the population (b). The solid lines are Student's $t$ distributions fitting to the data with parameters $v=1.51$ and $L=16.02$ for the primary industry and $v=1.25$ and $L=86.01$ for the population, respectively. The dot-dashed lines are the fits to Eq.~(\ref{Eq:PNAS:PDF}) from Ref.~\cite{Fu-Pammolli-Buldyrev-Riccaboni-Matia-Yamasaki-Stanley-2005-PNAS}.}
\end{figure}

Figure~\ref{Fig:PDF} illustrates the empirical probability distribution functions $f(r)$ of the standardized annual growth rates of the primary industry and the population of 2079 counties in mainland China. We find that Student's $t$
distribution fits the data well, which reads
\begin{equation}
    f(r)=\frac{v^{\frac{1}{2}v}}{B\left(\frac{1}{2},\frac{1}{2}v\right)}\left[v+L(r-m)^{2}\right]^{-\frac{1}{2}(v+1)}\sqrt{L}~,
    \label{Eq:t_distribution}
\end{equation}
where $v$ is the degrees of freedom parameter, $m$ is the location parameter and $L$ is the scale parameter, $B(a,b) = \Gamma(a)\Gamma(b)/\Gamma(a+b)$ is the Beta function, where $\Gamma(\cdot)$ is the Gamma function. Since the growth rates are standardized, we have $m=0$. Using the least-squares fitting method, we obtain $v=1.51$ and $L=16.02$ for the primary industry growth rate and $v=1.25$ and $L=86.01$ for the population growth rate.
We note that Student's $t$ distribution is in essence the $q$-Gaussian distribution \cite{Tsallis-1988-JSP}, which finds its applications in financial returns \cite{Tsallis-Anteneodo-Borland-Osorio-2003-PA,Osorio-Borland-Tsallis-2004,Queiros-Moyano-deSouza-Tsallis-2007-EPJB,Rak-Drozdz-Kwapien-2007-PA,Drozdz-Forczek-Kwapien-Oswicimka-Rak-2007-PA,Gu-Chen-Zhou-2008a-PA,Gu-Zhou-2009-EPJB}.

In order to test whether the annual growth rates $r$ of the primary industry and the population are drawn from Student's $t$ distributions, we use the Kolmogorov-Smirnov (KS) test, where the KS statistic is defined as
\begin{equation}
  KS=\max|F-P|,
  \label{Eq:KS}
\end{equation}
where $F$ is the cumulative distribution of the best fit and $P$ is the cumulative distribution of the empirical or synthetic data. We generate $1000$ sequences from the fitted Student distribution and calculate the KS statistic for each sequence. The $p$-value is then calculated as the percentage that the KS statistic of the synthetic sequences is greater than that of the original sequence. We find that the $p$-values are 0.94 for the primary industry data and 0.80 for the population data, which implies that the hypothesis that the annual growth rates $r$ for both the primary industry and the population follow the $t$ distribution cannot be rejected.

For comparison, we also adopt the theoretical form derived for firm growth \cite{Fu-Pammolli-Buldyrev-Riccaboni-Matia-Yamasaki-Stanley-2005-PNAS}
\begin{equation}
  f(r)=\frac{2V}{\sqrt{r^{2}+2V}\left(|r|+\sqrt{r^{2}+2V}\right)}.
  \label{Eq:PNAS:PDF}
\end{equation}
The resulting fits are also illustrated in Fig.~\ref{Fig:PDF}. The KS tests unveil that the $p$-values are 0.90 and 0.00 for the primary industry and the population, respectively. Therefore, the distribution (\ref{Eq:PNAS:PDF}) can be used to model the growth rates of the primary industry but not for the population. This finding is rational since the growth of primary industry is expected to have a more similar mechanism as firms than the growth of population.

We also use an alternative goodness-of-fit measure based on the CvM statistic \cite{Pearson-Stephens-1962-Bm,Stephens-1964-Bm,Stephens-1970-JRSSB}, which is defined as:
\begin{equation}
    W_{2}=\int_{-\infty}^{+\infty}[F_{N}(x)-F(x)]^{2}dF(x),
    \label{Eq:CvM}
\end{equation}
where $F_{N}$ is the empirical distribution function of the sample and $F(x)$ is the a specified continuous distribution. We find qualitatively the same results as the KS test. Therefore, we find that the Student distribution is a better model for the growth rates of the primary industry and the population. We note that it is impossible to provide a proof for a specific distribution \cite{Clauset-Shalizi-Newman-2009-SIAMR}. Additionally, a reasonable quantification of the distribution tail behavior is difficult for such a limited data set.

\subsection{Conditional distributions}
\label{S2:PDF:Cond}

\begin{figure}[htb]
\centering
\includegraphics[width=8cm]{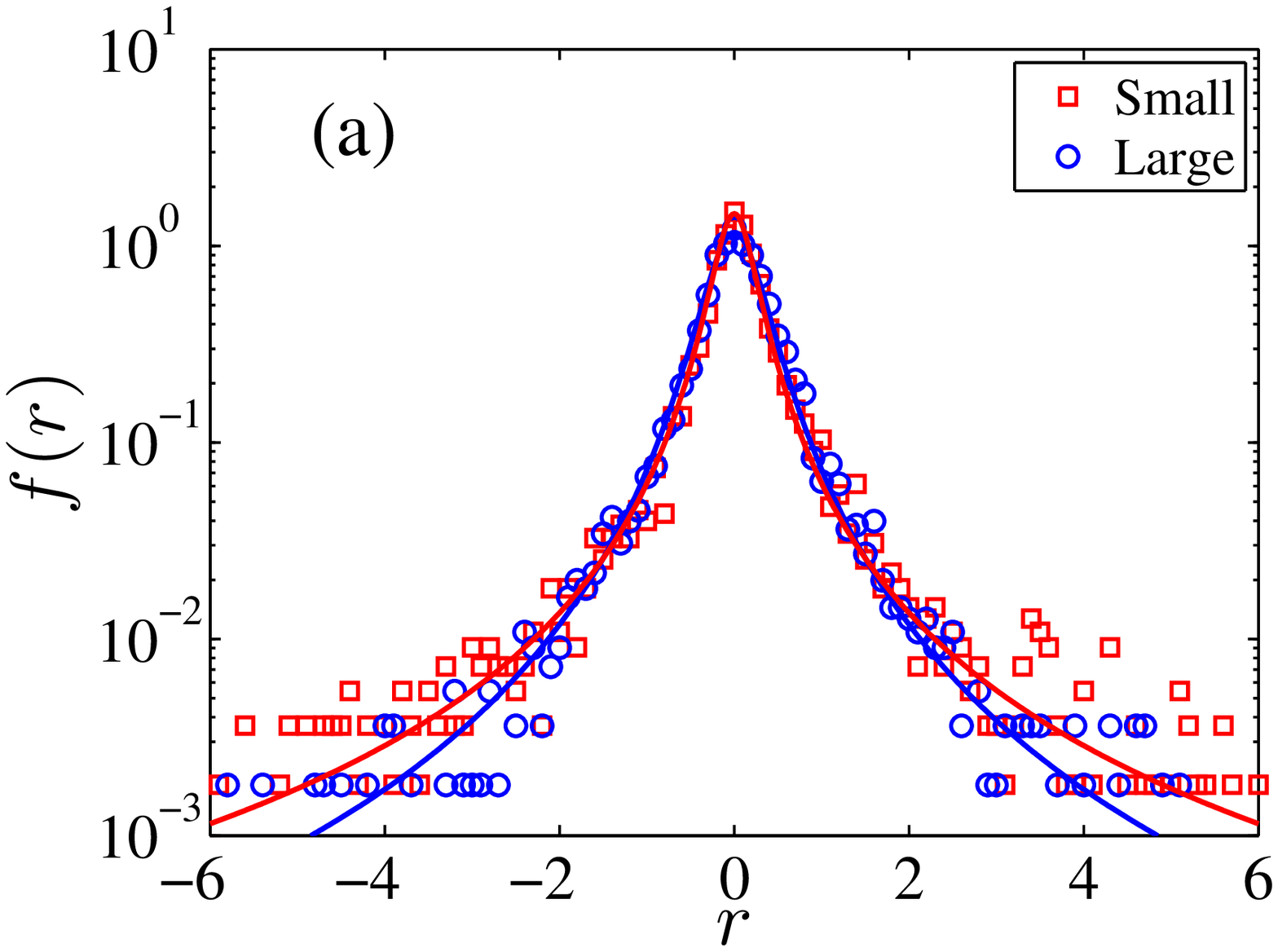}
\includegraphics[width=8cm]{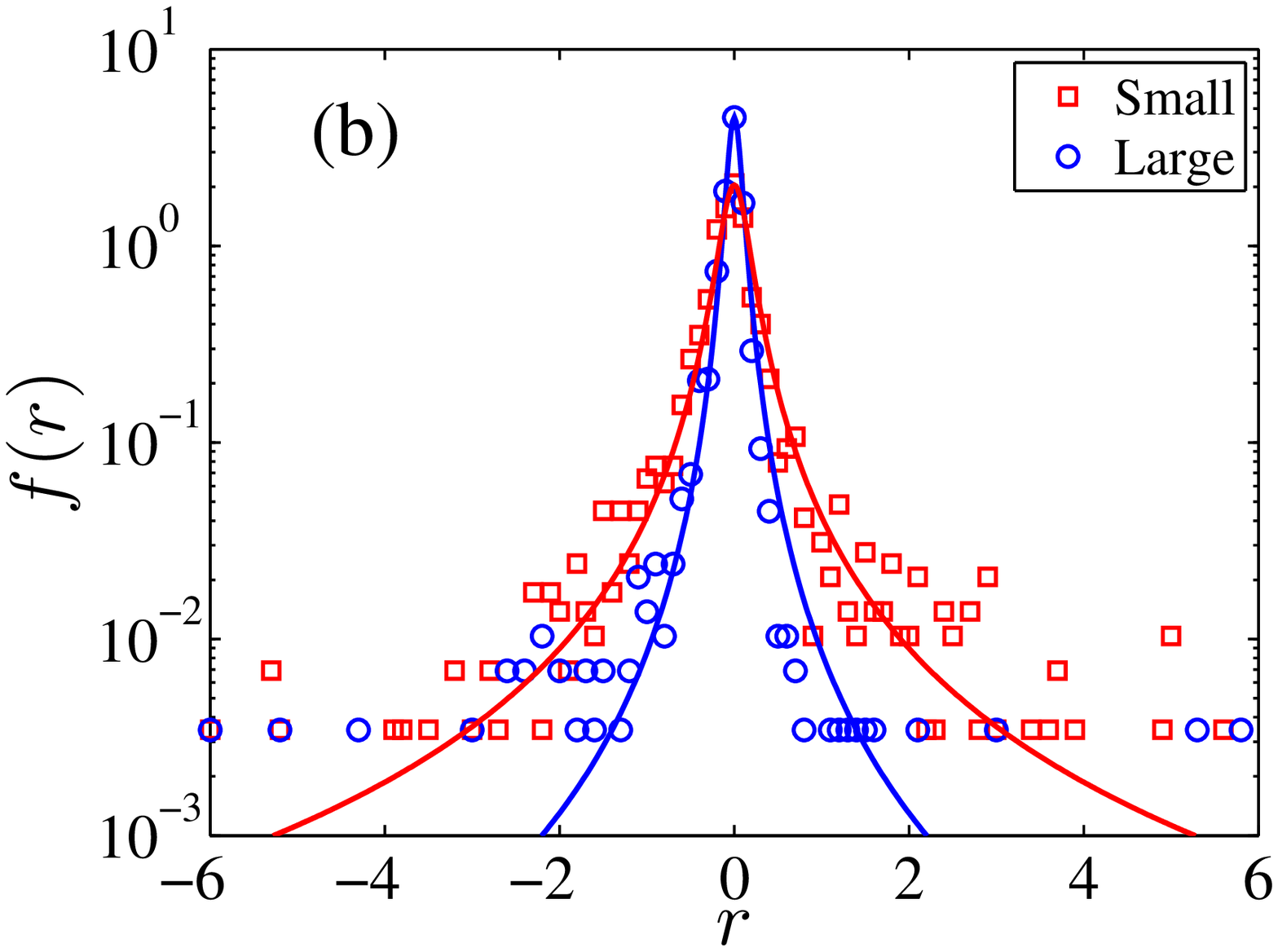}
\caption{\label{Fig:PDF:Cond} (Color online) (a) Plot of empirical conditional probability density function $f(r|S)$ for two bins of primary industry. The solid lines are the $t$ distributions fitting to the data in each bin with the parameters $v=1.26$ and $L=19.33$ for the small bin of primary industry and $v=1.85$ and $L=11.16$ for the large one. (b) Plot of empirical conditional probability density function $f(r|S)$ for two bins of population. The solid lines are the $t$ distributions fitting to the data in each bin with the parameters $v=1.27$ and $L=37.78$ for the small bin of primary industry and $v=1.73$ and $L=168.55$ for the large one.}
\end{figure}

We further investigate whether the distributions of the growth rates are dependent on the initial values of the primary industry and the population. The entire growth rates of the primary industry are first arranged in an ascending order then grouped into two bins with identical number of data, that is, $1.6\times10^{6}<S\leqslant4\times10^{8}$ (small bin) and $4\times10^{8}<S<3.6\times10^{9}$ (large bin), where $S$ is the value of primary industry. For each bin we calculate the conditional probability density function $f(r|S)$. The results are shown in Figure~\ref{Fig:PDF:Cond}(a). We find that the data in each bin follows the $t$ distribution with different parameters. Using the least-squares fitting method, we have $v=1.26$ and $L=19.33$ for the small bin of primary industry and $v=1.85$ and $L=11.16$ for the large bin. Similarly, we divide the population growth rates into two bins, $0.95\times10^{4}<S\leqslant5\times10^{5}$ (small bin) and $5\times10^{5}<S\leqslant2.5\times10^{7}$ (large bin). We find that the data in the two bins obey the $t$ distribution, which are presented in Figure~\ref{Fig:PDF:Cond}(b), with the parameters $v=1.27$ and $L=37.78$ for the small bin and $v=1.73$ and $L=168.55$ for the large bin.

\subsection{Power-law tails}
\label{S2:PDF:tails}

\begin{figure}[htb]
\centering
\includegraphics[width=5.3cm]{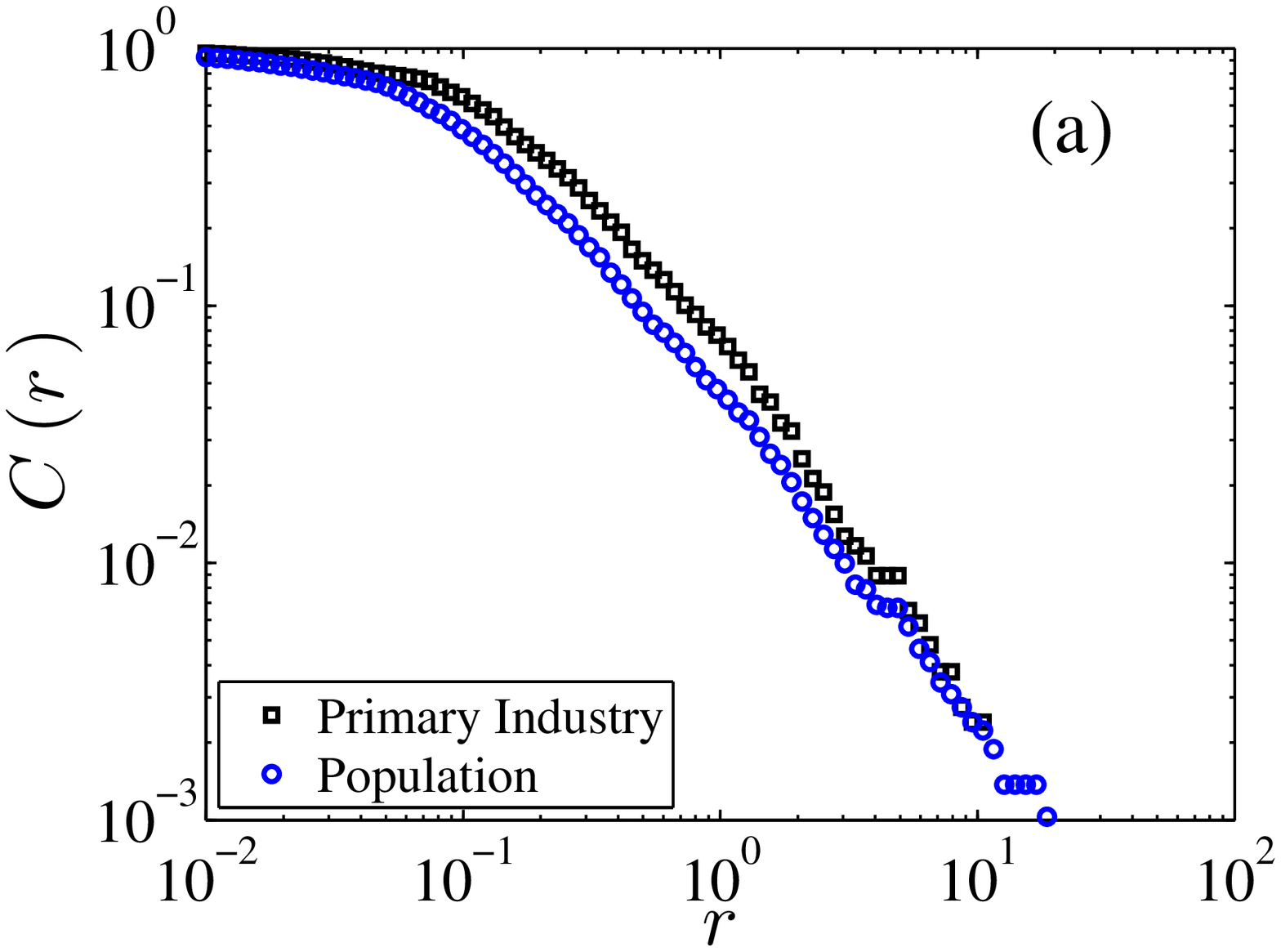}
\includegraphics[width=5.3cm]{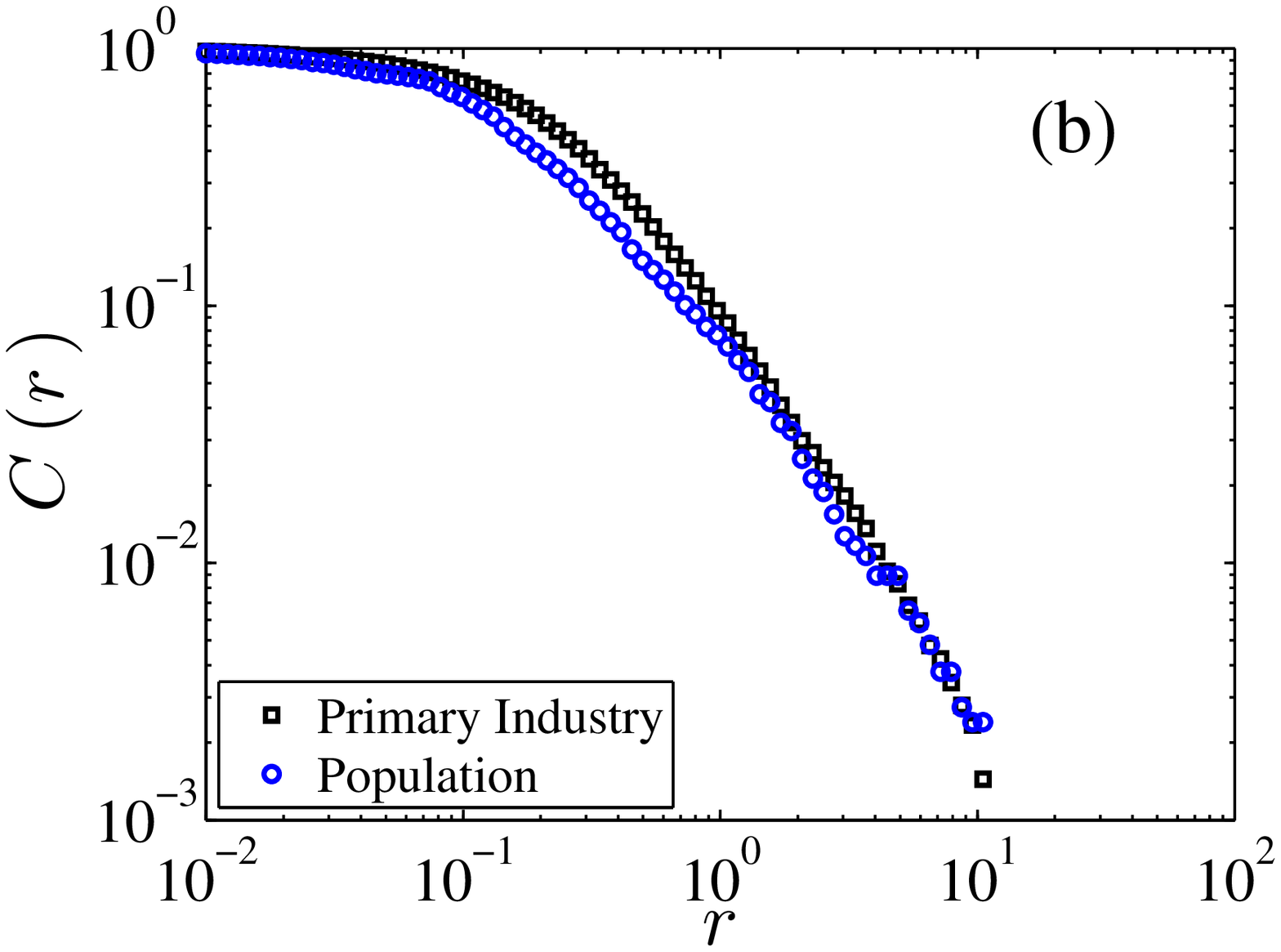}
\includegraphics[width=5.3cm]{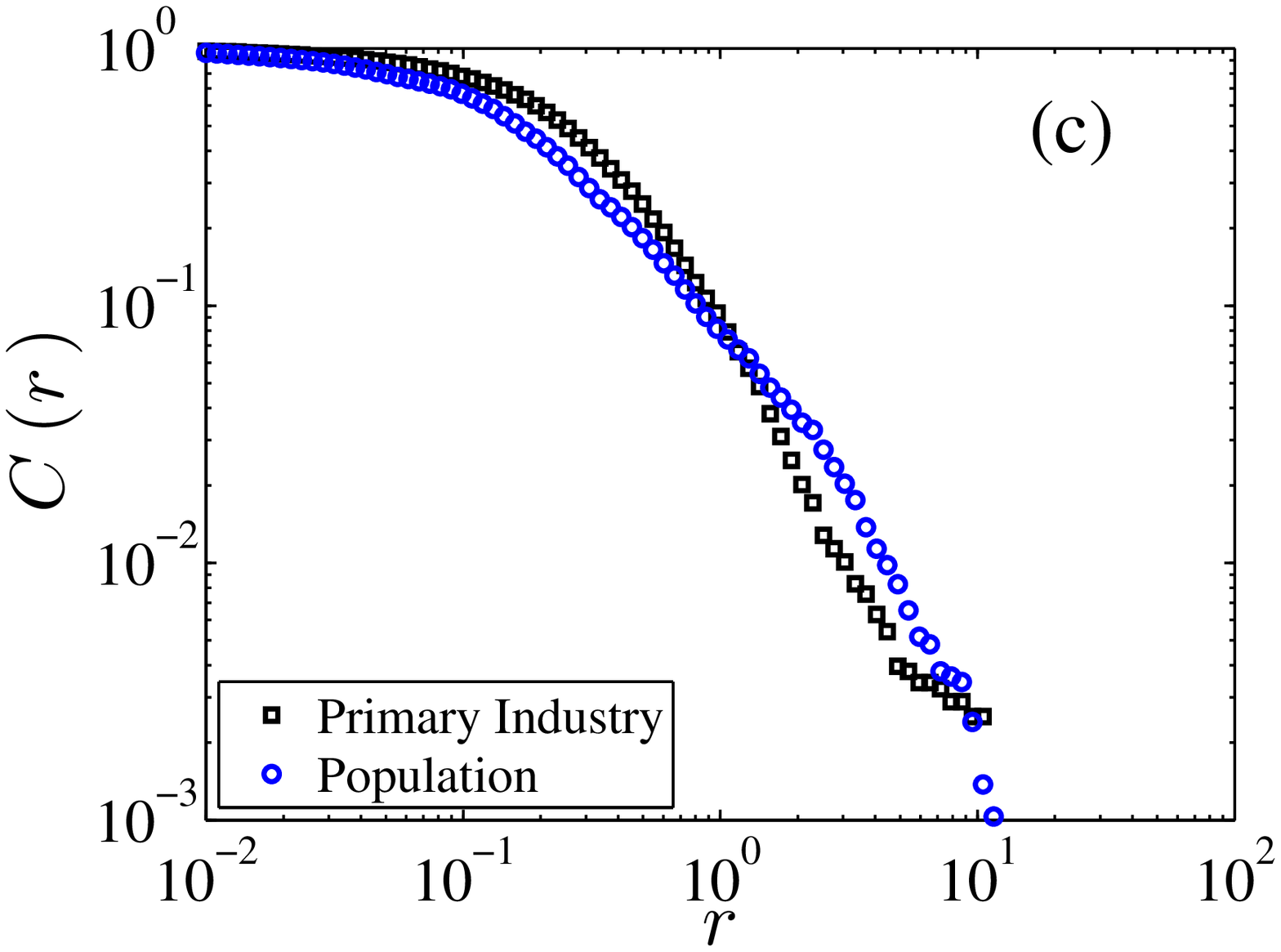}
\caption{\label{Fig:PDF:tails} Empirical complementary cumulative distribution functions $C(r)$ for the primary industry and the population: (a) all sizes, (b) small sizes, and (c) large sizes.}
\end{figure}

In the preceding two subsections, we have seen that the growth rates have power-law tails for both the primary industry and the population. Here we further investigate the tails of the complementary cumulative distributions $C(r)=\Pr(\geqslant{r})$ of the whole data sets in Section \ref{S2:PDF:unCond} and the subgroups in Section \ref{S2:PDF:Cond}. The empirical complementary cumulative distributions are illustrated in Fig.~\ref{Fig:PDF:tails}. Eye-balling unveils that there are power-law tails in these empirical distributions
\begin{equation}
 C(r) \sim r^{-\alpha},
 \label{Eq:Cr:PL}
\end{equation}
where the tail exponent $\alpha$ can be determined in a statistical way.

In order to determine the tail exponents, we adopt a nonparametric maximum likelihood estimation method proposed by Clauset, Shalizi and Newman (CSN) \cite{Clauset-Shalizi-Newman-2009-SIAMR}. The CSN method fits the power-law distribution to the data, along with the goodness-of-fit based approach to estimating the lower cutoff $x_{\min}$ for the scaling region. Having $x_{\min}$ enables us to determine the percentage of the data points in the tails that are included in the fitting. The $p$-values are also calculated based on the Kolmogorov-Smirnov test for the power-law model (\ref{Eq:Cr:PL}). The results are presented in Table \ref{Tb:TailExponents}. We find that all the tail exponents lie in the interval $(1,2)$, which means that the distributions are within the L\'evy regime \cite{Mantegna-Stanley-2000}. For comparison, we also list in Table \ref{Tb:TailExponents} the values of $v$ in the $t$ distribution. It is found that these values are also less than 2 and the two sets of exponents estimated from the two methods are consistent with each other. This finding implies that the growth dynamics of the primary industry and the population studied in this work fall in to a distinct regime that is different from the growth dynamics of firms where the tail exponent is $\zeta-1\approx2$ \cite{Fu-Pammolli-Buldyrev-Riccaboni-Matia-Yamasaki-Stanley-2005-PNAS,Pammolli-Fu-Buldyrev-Riccaboni-Matia-Yamasaki-Stanley-2007-EPJB}.

\begin{table}[htp]
 \centering
 \caption{Estimation of the tail exponents using the nonparametric approach of Clauset, Shalizi and Newman \cite{Clauset-Shalizi-Newman-2009-SIAMR} with comparison to the fitted values using the $t$ distribution (\ref{Eq:t_distribution}). The second row of the CSN method gives the percentages of the data points included in the fitting.}
 \medskip
 \label{Tb:TailExponents}
 \begin{tabular}{cccccccccccc}
 \hline \hline
    \multirow{4}*[3.5mm]{Method} & \multirow{4}*[3.5mm]{Parameters} && \multicolumn{3}{c}{Primary Industry} && \multicolumn{3}{c}{Population} \\
    \cline{4-6} \cline{8-10}
    &&& All & Small & Large  && All & Small & Large \\
    \hline
                          & $x_{\min}$ &&  1.10    &  0.38    &   0.98    &&   0.26     &  0.26    &   0.39   \\
    CSN      & pctg\%     &&  8.08    & 30.24    &   9.10    &&  20.22     & 31.29    &  19.94   \\
                          & $\alpha$   &&  1.58    &  1.20    &   1.87    &&   1.16     &  1.15    &   1.45   \\\hline
\multirow{2}*[0.1mm]{$t$ distribution} & $v$        &&  1.51    &  1.26    &   1.85    &&   1.25     &  1.27    &   1.73   \\
                          & $L$        && 16.02    & 19.33    &  11.16    &&  86.01     & 37.78    & 168.55   \\
 \hline\hline
 \end{tabular}
\end{table}

%
%

\section{Size-dependent standard deviation}
\label{S1:Std:Size}

\begin{figure}[htb]
\centering
\includegraphics[width=8cm]{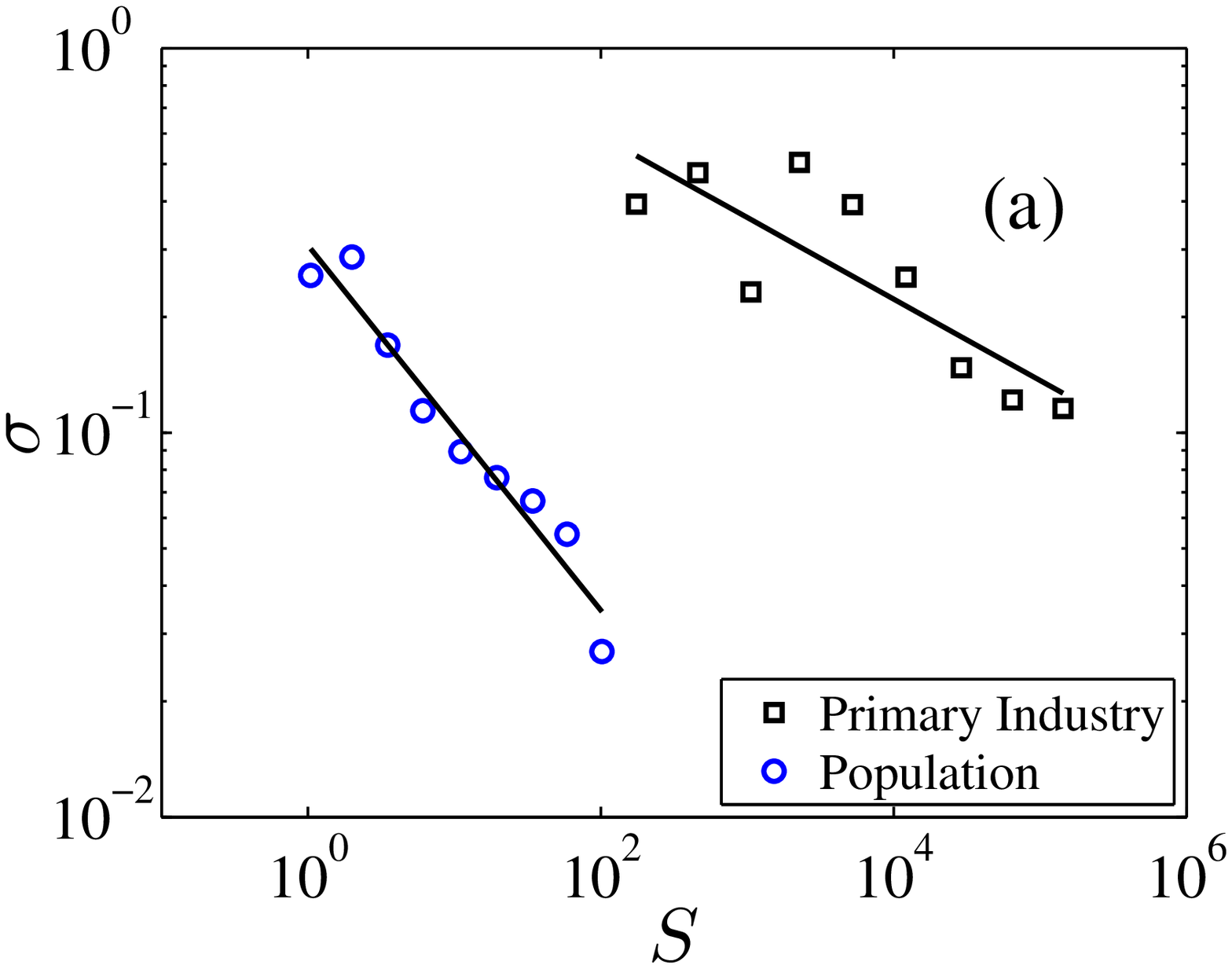}
\includegraphics[width=8cm]{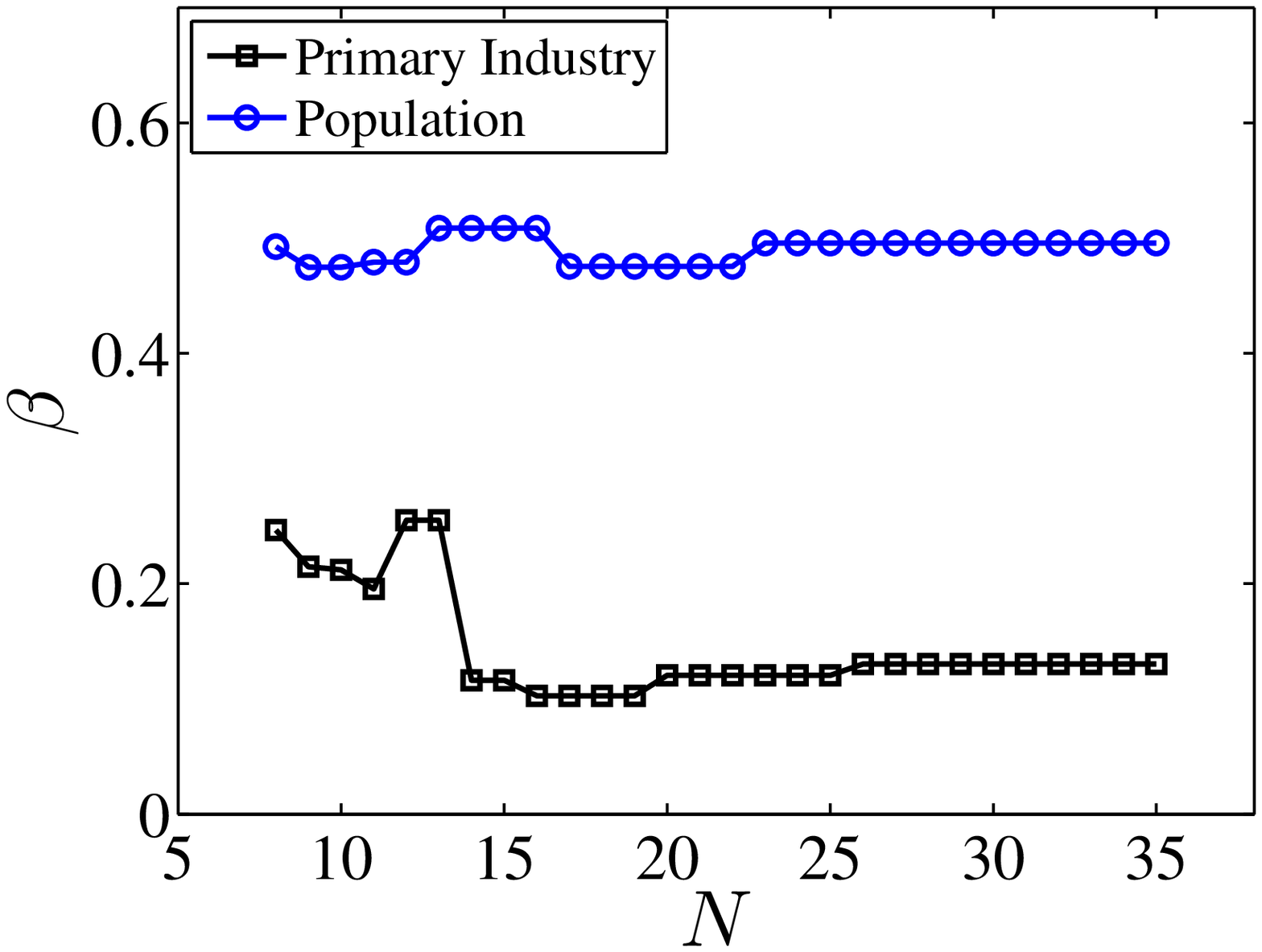}
\caption{\label{Fig:Sigma:S} (a) Dependence of the standard deviation of the growth rate with respect to the initial size in log-log coordinates for both the primary industry and the population. The solid lines are the power-law fits to the data with the scaling exponent $\beta=0.21\pm0.06$ for the primary industry and $\beta=0.47\pm0.03$ for the population. (b) Dependence of the scaling exponent $\beta$ as a function of the number of bins.}
\end{figure}

In this section, we analyze the relationship between the standard deviation $\sigma$ of growth rate and the initial size $S$ used to calculate the growth rate. Unfortunately, as shown in Section \ref{S1:PDF}, the degree of freedom $v$ in Student's $t$ distribution is less than 2. The theoretical consequence is that the variance of the growth rates diverges. Nevertheless, we calculate the sample variance $\sigma^2$ regardless of its asymptotic divergence. The data points are rearranged so that the sizes $S$ are in an ascending order. We then partition the data points into $N$ groups with identical number of data. For each group, we calculate the average size $S$ and the empirical standard deviation $\sigma$.
Fig.~\ref{Fig:Sigma:S}a plots $\sigma$ against $S$ for $N=9$ for both the primary industry and the population. Power-law relations are observed for both cases, as expressed in Eq.~(\ref{Eq:sigma:r:S}). We find that $\beta=0.21\pm0.06$ for the primary industry and $\beta=0.47\pm0.03$ for the population.

We change the number $N$ of bins to repeat the same analysis. The value of $N$ varies from 8 to 35 with a step of 1. Similar power-law relations are found between the standard deviation $\sigma$ and the size $S$ for all the $N$ values. Fig.~\ref{Fig:Sigma:S}b illustrates the dependence of the estimated exponent $\beta$ as a function of $N$. For the primary industry, $\beta$ fluctuates between 0.1 and 0.25, which is consistent with previous results on the GDP and firm sales. For the population, the exponent $\beta$ is approximately equal to 0.5 with slight fluctuations within $(0.47,0.51)$. The case of the population growth is very interesting, which gives a nice example for the dynamic growth model of a complex organization containing subunits without interactions \cite{Amaral-Buldyrev-Havlin-Salinger-Stanley-1998-PRL,Lee-Amaral-Canning-Meyer-Stanley-1998-PRL,Amaral-Gopikrishnan-Plerou-Stanley-2001-PA}. Indeed, we can divide each county into many subunits of identical size. These subunits are not necessary to be administrative districts or villages. It is rational that there is no interaction among different subunits in the population growth. It follows immediately that $\beta=0.5$ \cite{Amaral-Buldyrev-Havlin-Salinger-Stanley-1998-PRL,Lee-Amaral-Canning-Meyer-Stanley-1998-PRL,Amaral-Gopikrishnan-Plerou-Stanley-2001-PA}. This argument can be modeled using the city clustering algorithm, which leads to the conclusion that the exponent $\beta$ is related to the long-range spatial correlations in population growth and $\beta=0.5$ if there are no spatial correlations \cite{Rozenfeld-Rybski-Andrade-Batty-Stanley-Makse-2008-PNAS}.

\section{Conclusion}
\label{S1:Conclusion}

In this work, we have investigated the growth dynamics of the primary industry and the population of 2079 counties in mainland China using the data from the China County Statistical Yearbooks from 2000 to 2006. We found that the annual growth rates are distributed according to Student's $t$ distribution with the degree of freedom less than 2 for both the primary industry and the population. When each sample is divided into to subgroups according to their initial size, the data for each subgroup also has a $t$ distribution. The power-law behavior in the tails of the distributions is further confirmed by a nonparametric approach based on the maximum likelihood estimation and the Kolmogorov-Smirnov test \cite{Clauset-Shalizi-Newman-2009-SIAMR}. The fact that tail exponent is less than 2 means that the annual growth rates fall in the L\'evy regime, which is different from that of firms. Therefore, the growth dynamics of the primary industry and the population at the county level in mainland China might be different from other complex organizations studied in the literature, especially by the Boston School.

We also investigated the relationship between the sample standard deviation of the growth rates and the initial size of the primary industry and the population, despite of the fact that the variances of the growth rates theoretically diverge. We observed power-law dependence between the two quantities. For the primary industry, the scaling exponent is less than 0.5, which is consistent with the fact that the economic systems between different counties have strong interactions. In contrast, for the population, no interactions between the growth dynamics of any two counties can be pinpointed and the scaling exponent is close to 0.5, as expected by theoretical analysis \cite{Amaral-Buldyrev-Havlin-Salinger-Stanley-1998-PRL,Lee-Amaral-Canning-Meyer-Stanley-1998-PRL,Amaral-Gopikrishnan-Plerou-Stanley-2001-PA}.

In summary, we have uncovered idiosyncratic properties in the growth dynamics of the primary industry and the population of a developing country at the county level. Further research should be done to understand the underlying microscopic mechanism causing such idiosyncracy. Many important questions are not clear that should be addressed. For instance, we do not know if the Chinese firms have similar growth dynamics as those in the developed countries. Such comparative studies will help us to better understand socio-economic systems.

\bigskip

{\textbf{Acknowledgments:}}

We are grateful to Diego Rybski for discussions. This work was partially supported by the Program for New Century Excellent Talents in University (NCET-07-0288).

\bibliography{E:/papers/Auxiliary/Bibliography}

\end{document}